\documentclass[aps,prb,reprint,superscriptaddress]{revtex4-2}
\usepackage{graphicx}
\usepackage{amsmath}
\usepackage{amssymb}
\usepackage{bm}
\usepackage{physics}
\usepackage{subfigure}
\usepackage[colorlinks=true,linkcolor=blue,urlcolor=blue,citecolor=blue]{hyperref}
\usepackage{times}
\usepackage{changes}
\begin{document}

\newcommand{\Hop}{\hat{H}}
\newcommand{\Himp}{\hat{H}_{\rm imp}}
\newcommand{\Hint}{\hat{H}_{\rm int}}
\newcommand{\Uimp}{\hat{U}_{\rm imp}}
\newcommand{\gHimp}{\mathcal{H}_{\rm imp}}
\newcommand{\gUimp}{\mathcal{U}_{\rm imp}}
\newcommand{\Hbath}{\hat{H}_{\rm bath}}
\newcommand{\Hhyb}{\hat{H}_{\rm hyb}}

\newcommand{\aop}{\hat{a}}
\newcommand{\adop}{\hat{a}^{\dagger}}
\newcommand{\sgp}{\hat{\sigma}^+}
\newcommand{\sgx}{\hat{\sigma}^x}
\newcommand{\sgy}{\hat{\sigma}^y}
\newcommand{\sgz}{\hat{\sigma}^z}
\newcommand{\nop}{\hat{n}}

\newcommand{\cop}{\hat{c}}
\newcommand{\cdop}{\hat{c}^{\dagger}}
\newcommand{\hc}{{\rm H.c.}}
\newcommand{\rhotot}{\hat{\rho}_{\mathrm{tot}}}
\newcommand{\rhoop}{\hat{\rho}}
\newcommand{\rhoimp}{\hat{\rho}_{\mathrm{imp}}}
\newcommand{\rhobath}{\hat{\rho}_{\mathrm{bath}}}
\newcommand{\Zimp}{Z_{{\rm imp}}}
\newcommand{\mea}{\mathcal{D}}
\newcommand{\gK}{\mathcal{K}}
\newcommand{\gI}{\mathcal{I}}
\newcommand{\gF}{\mathcal{F}}
\newcommand{\bolda}{\bm{a}}
\newcommand{\boldabar}{\bar{\bm{a}}}
\newcommand{\abar}{\bar{a}}
\newcommand{\im}{{\rm i}}
\newcommand{\contour}{\mathcal{C}}
\newcommand{\gA}{\mathcal{A}}
\newcommand{\gB}{\mathcal{B}}
\newcommand{\gM}{\mathcal{M}}
\newcommand{\boldeta}{\bm{\eta}}
\newcommand{\boldetabar}{\bar{\bm{\eta}}}
\newcommand{\etabar}{\bar{\eta}}
\newcommand{\parity}{\mathcal{P}}
\newcommand{\current}{\mathcal{J}}
\newcommand{\pronyerror}{\varsigma_p}

\newcommand{\WI}{{\rm W}^I}
\newcommand{\WII}{{\rm W}^{II}}

\newcommand{\EqDef}{\stackrel{\mathrm{def}}{=}}

\newcommand{\rem}[1]{{\color{red}{\sout{#1}}}}
\newcommand{\gcc}[1]{{\color{black}#1}}

\definechangesauthor[name=RF,color=red]{RF}

\title{Infinite Grassmann Time-Evolving Matrix Product Operator Method in the Steady State}


\author{Chu Guo}
\email{guochu604b@gmail.com}
\affiliation{Key Laboratory of Low-Dimensional Quantum Structures and Quantum Control of Ministry of Education, Department of Physics and Synergetic Innovation Center for Quantum Effects and Applications, Hunan Normal University, Changsha 410081, China}

\author{Ruofan Chen}
\email{physcrf@sicnu.edu.cn}
\affiliation{College of Physics and Electronic Engineering, and Center for Computational Sciences, Sichuan Normal University, Chengdu 610068, China}

\date{\today}

\begin{abstract}
We present an infinite Grassmann time-evolving matrix product operator method for quantum impurity problems, which directly works in the steady state. The method embraces the well-established infinite matrix product state algorithms with the recently developed GTEMPO method, and benefits from both sides: it obtains real-time Green's functions without sampling noises and bath discretization error, it is applicable for any temperature without the sign problem, its computational cost is independent of the transient dynamics and does not scale with the number of baths. 
We benchmark the method on the finite-temperature equilibrium Green's function in the noninteracting limit against exact solutions and in the single-orbital Anderson impurity model against GTEMPO calculations. We also study the zero-temperature non-equilibrium steady state of an impurity coupled to two baths with a voltage bias, obtaining consistent particle currents with existing calculations. The method is ideal for studying steady-state quantum transport, and can be readily used as an efficient real-time impurity solver in the dynamical mean field theory and its non-equilibrium extension.
\end{abstract}
\maketitle


\section{Introduction}

Understanding non-equilibrium and open quantum phenomena is one of the major pursuits since the born of quantum physics. A prototypical microscopic model to describe these phenomena is the Anderson impurity model (AIM), where an impurity is coupled to one or several continuous, noninteracting baths of itinerant electrons~\cite{anderson1961-localized}. 
By imposing a temperature or a voltage bias among the baths, the whole system of the impurity plus baths will be driven to a non-equilibrium steady state (NESS).
The interplay between the strong local Coulomb interaction, the non-equilibrium driving and the dissipative effects of the bath could lead to very rich physical phenomena in the NESS~\cite{hettler1998-nonequilibrium,muhlbacher2011-anderson,becker2012-non,cohen2013-numerically,AokiWerner2014,dorda2015-auxiliary,fugger2020-nonequilibrium}.

A variety of numerical methods have been developed to study steady-state quantum transport~\cite{RoschWolfle2003,Kehrein2005,JakobsSchoeller2007,BoulatSchmitteckert2008,Anders2008,MeisnerDagotto2009,EckelEgger2010,PhilippMillis2010,MikhailHerbert2012,SergeyMilena2013,CohenMillis2014,ReininghausSchoeller2014,AntipovGull2016,ErpenbeckThoss2018,SchwarzWeichselbaum2018,LotemGoldstein2020,BertrandWaintal2019,ErpenbeckCohen2023}, which could provide fairly accurate solutions in specific regimes. However, up to date there is no single method which could provide generally reliable and efficient solutions, as similar to the role played by the continuous-time Quantum Monte Carlo (CTQMC) methods in solving the equilibrium AIM in the imaginary-time axis~\cite{GullWerner2011,RubtsovLichtenstein2005,GullTroyer2008,WernerMillis2006b,WernerMillis2006,ShinaokaWerner2017,EidelsteinCohen2020}.

The Grassmann time-evolving matrix product operator (GTEMPO) method, recently developed by us, is a promising candidate of this kind to solve the non-equilibrium quantum transport problem of the AIM~\cite{ChenGuo2024a}: it treats the bath exactly and obtains results in the real-time axis without sampling noises; it is applicable for any temperature without the sign problem; most remarkably, its computational cost is independent of the number of baths, a feature that is ideal for studying quantum transport and is missing in most existing alternatives. \gcc{The GTEMPO method is also an extension of the time-evolving matrix product operator method~\cite{StrathearnLovett2018}, which is the state-of-the-art for bosonic impurity problems~\cite{LeggettZwerger1987,joergensen2019-exploiting,popovic2021-quantum,fux2021-efficient,gribben2021-using,otterpohl2022-hidden,gribben2022-exact,chen2023-heat,kamar2023spin}, to fermionic impurity problems.}

The power of GTEMPO can be understood through its formalism.
Roughly speaking, its idea is somewhere in between the CTQMC methods and the conventional wave-function based methods where the impurity-bath state is parametrized by some wave function ansatz~\cite{CaffarelKrauth1994,KochGunnarsson2008,GranathStrand2012,LuHaverkort2014,ZaeraLin2020,HeLu2014,HeLu2015,WolfSchollwock2014b,GanahlEvertz2014,GanahlVerstraete2015,WolfSchollwock2015,GarciaRozenberg2004,NishimotoJeckelmann2006,WeichselbaumDelft2009,BauernfeindEvertz2017,LuHaverkort2019,WernerArrigoni2023,KohnSantoro2021,KohnSantoro2022}. Similar to CTQMC, it integrates out the bath exactly. However, instead of sampling from the perturbative expansion of the path integral (PI), GTEMPO directly represents the integrand of the PI as a Grassmann matrix product state (GMPS) non-perturbatively.
Similar to the wave-function based methods, GTEMPO uses GMPS as its parametric ansatz, but only in the temporal domain for the multi-time impurity degrees of freedom. The computational cost of GTEMPO roughly scales as $N^2\chi^3$ for the single-orbital AIM, with $N$ the discrete time steps and $\chi$ the bond dimension of the GMPS. Importantly, for commonly used bath spectrum density, it is observed that $\chi\approx 100$ could already give very accurate results~\cite{ChenGuo2024a,ChenGuo2024b,GuoChen2024d}, which underlies the efficiency of GTEMPO.
However, to study steady-state quantum transport with GTEMPO, one needs to overcome the (usually very long) transient dynamics to reach the NESS first~\cite{ChenGuo2024c}, as similar to the wave-function based methods~\cite{BauerTroyer2016,KohnSantoro2021,KohnSantoro2022}, which sets a very large prefactor $N$ in the computational cost that could greatly hinder the efficiency.
Here we also note the closely related tensor network IF method~\cite{ThoennissAbanin2023a,ThoennissAbanin2023b}, which represents the Feynman-Vernon IF per spin per bath as a fermionic matrix product state (MPS) in the Fock state basis, therefore its computational cost scales exponentially with the number of baths.



In this work we propose an infinite GTEMPO (referred to as iGTEMPO afterwards) method which directly targets at the steady state, eliminating the need for transient dynamics. The key insight is that 
in the infinite-time limit the memory of the initial state is lost, which is a situation that closely resembles an infinite many-body wave function where the boundary condition becomes irrelevant if one is only interested in the bulk. Therefore similar to the spatial case, 
one could use infinite GMPSs instead of finite ones to represent the multi-time impurity steady states, in which only the tensors within a single time step need to be stored and manipulated. 
The iGTEMPO method seamlessly integrates the well-established infinite MPS techniques into the GTEMPO formalism. 
We benchmark its accuracy on the finite-temperature equilibrium Green's function in the noninteracting limit against exact solutions, and in the single-orbital AIM against GTEMPO calculations. We also apply it to study the NESS of an impurity that is coupled to two zero-temperature baths with a voltage bias, where the obtained steady-state particle currents are consistent with existing calculations.


\section{The GTEMPO Method}

Before we introduce the iGTEMPO method, we first give an elementary review of the GTEMPO method in the real-time axis as they share most of the techniques in common. We will mainly focus on the single-orbital AIM (although both methods are directly applicable for general impurity models), where the impurity may be coupled to one bath (the equilibrium setup) or two baths (the non-equilibrium setup). The Hamiltonian is denoted as $\Hop = \Himp + \Hint$, where
\begin{equation}
  \Himp = (\epsilon_d-U/2)\sum_{\sigma}\adop_{\sigma}\aop_{\sigma} + U \adop_{\uparrow}\aop_{\uparrow}\adop_{\downarrow}\aop_{\downarrow}
\end{equation}
is the impurity Hamiltonian with $\epsilon_d$ the on-site energy and $U$ the local Coulomb interaction,
\begin{equation}
  \Hint = \sum_{\nu, k, \sigma}\epsilon_k \cdop_{\nu,k, \sigma}\cop_{\nu,k, \sigma} + \sum_{\nu,k, \sigma}\left(V_{\nu,k} \adop_{\sigma}\cop_{\nu,k, \sigma} + \hc \right)
\end{equation}
contains the free bath Hamiltonians and the coupling between the impurity and the baths, characterized by the band energy $\epsilon_k$ and the coupling strength $V_{\nu,k}$ respectively. Here we have used $\sigma \in \{\uparrow, \downarrow\}$ for spin indices, and $\nu$ as the bath label.

The GTEMPO method is essentially a translation of the Grassmann PI into efficient GMPS operations. For real-time dynamics starting from a non-correlated impurity-bath initial state $\rhoop(0) = \rhoimp \otimes	\rhobath^{\rm th} $, where $\rhoimp$ is some impurity initial state and $\rhobath^{\rm th}$ is the bath equilibrium state, the corresponding path integral of the impurity partition function $\Zimp(t) = \Tr \rhoop(t) / \Tr \rhobath^{\rm th} $ can be written as 
\begin{align}\label{eq:PI}
\Zimp(t) = \int \mathcal{D}[\boldabar,\bolda] \gK\left[\boldabar, \bolda \right]\prod_{\sigma}\gI_{\sigma}\left[\boldabar_{\sigma}, \bolda_{\sigma}\right].
\end{align}
Here $\boldabar_{\sigma} = \{\abar_{\sigma}(\tau)\}$, $\bolda_{\sigma} = \{a_{\sigma}(\tau)\}$ are Grassmann trajectories \cite{negele1998-quantum} on the Keldysh contour \cite{lifshitz1980-statistical,wang2013-nonequilibrium,kamenev2009-keldysh}, and $\boldabar=\{\boldabar_{\uparrow},\boldabar_{\downarrow}\},\bolda=\{\bolda_{\uparrow},\bolda_{\downarrow}\}$. The measure is
\begin{equation}
\mathcal{D}[\boldabar,\bolda]=\prod_{\sigma,\tau}d\abar_{\sigma}(\tau)d a_{\sigma}(\tau)
e^{-\abar_{\sigma}(\tau)a_{\sigma}(\tau)}.
\end{equation}
$\gI_{\sigma}$ is the Feynman-Vernon influence functional (IF)~\cite{FeynmanVernon1963}:
\gcc{
\begin{align}\label{eq:I}
\gI_{\sigma}[\boldabar_{\sigma}, \bolda_{\sigma}] = e^{-\int_{\contour} d t' \int_{\contour}d t'' \abar_{\sigma}(t') \Delta(t', t'') a_{\sigma}(t'') },
\end{align}
where $\contour$ is the Keldysh contour, and the hybridization function $\Delta(t', t'') = \sum_{\nu} \Delta_{\nu}(t', t'')$, with $\Delta_{\nu}(t', t'')$ being the hybridization function of bath $\nu$ (noticing that $\nu$ does not appear as a subscript of $\gI$), which is calculated by 
\begin{align}
\Delta_{\nu}(t', t'') = \mathcal{P}_{t't''}\int d\omega J_{\nu}(\omega) D_{\nu,\omega}(t, t'').
\end{align}
Here $J_{\nu}(\omega) = \sum_k V_{\nu,k}^2 \delta(\omega - \omega_k)$ is the $\nu$th bath spectrum density, $D_{\nu,\omega}(t', t'') = \langle T_{\contour}\cop_{\nu, \omega}(t')\cdop_{\nu, \omega}(t'') \rangle_0$ is the free contour-ordered Green’s function of the $\nu$th bath, $\mathcal{P}_{t't''}=1$ if $t', t''$ are on same Keldysh branch and $-1$ otherwise~\cite{ChenGuo2024a}.
}

After discretization using the quasi-adiabatic propagator path integral (QuaPI) method~\cite{makarov1994-path,makri1995-numerical,ChenGuo2024a} with a time step size $\delta t$, $\gI_{\sigma}$ can be written as (up to first-order time discretization error)
\begin{align}\label{eq:gI}
\gI_{\sigma} \approx e^{\gF_{\sigma}}, \gF_{\sigma} \EqDef -\sum_{\zeta,\zeta'=\pm}\sum_{j,k=1}^N\abar^{\zeta}_{\sigma,j}\Delta_{j,k}^{\zeta \zeta'}a^{\zeta'}_{\sigma,k},
\end{align}
where $N = t / \delta t$ is the total number of discrete time steps, $\zeta, \zeta'$ label the forward ($+$) and backward ($-$) Keldysh branches, \gcc{and $\Delta_{j,k}^{\zeta \zeta'}$ is the discretized hybridization function}.
$\gK$ encodes the bare impurity dynamics that is only determined by $\rhoimp$ and $\Himp$. After discretization, $\gK$ can be written as
\begin{align}\label{eq:gK}
    \mathcal{K}=&
    \langle -\bolda_0 \vert \bolda_{N}^+ \rangle \cdots
    \mel*{\bolda_2^+}{\Uimp}{\bolda_1^+} \mel*{\bolda_{1}^+}{\rhoimp}{\bolda_{1}^-}\times \nonumber \\
    & \mel*{\bolda_1^-}{\Uimp^{\dagger}}{\bolda_{2}^-}\times\cdots \times\langle \bolda_{N}^- \vert \bolda_0 \rangle,
\end{align}
where $\bolda_k^{\pm}=\{a_{\uparrow,k}^{\pm},a_{\downarrow,k}^{\pm}\}$ and $\Uimp = e^{-\im \Himp\delta t}$, $\mel*{\bolda_{1}^+}{\rhoimp}{\bolda_{1}^-}$ imposes the initial condition. The first and last terms on the rhs (with the auxiliary Grassmann variables $\bolda_0 = \{a_{\uparrow,0}, a_{\downarrow,0}\}$) impose the boundary condition, e.g., the final trace operation in the impurity partition function.

The discretized $\gK$ and $\gI_{\sigma}$ are Grassmann tensors. They can be multiplied together using Grassmann tensor multiplications~\cite{ChenGuo2024a} to obtain $\gA\left[\boldabar, \bolda \right] =  \gK\left[\boldabar, \bolda \right]\prod_{\sigma}\gI_{\sigma}\left[\boldabar_{\sigma}, \bolda_{\sigma}\right]$ as a single Grassmann tensor, which is the multi-time impurity state that encodes the whole information of the impurity dynamics, and is referred to as the augmented density tensor (ADT). In GTEMPO, one represents each $\gK$ and $\gI_{\sigma}$ as a GMPS, and then multiplies them together to obtain $\gA$ as a GMPS (this multiplication is only performed on the fly using a zip-up algorithm for efficiency~\cite{ChenGuo2024a}). Based on $\gA$ one can calculate any multi-time impurity correlations following the standard path integral formalism. For example, the greater and lesser Green's functions between two time steps $j$ and $0$ can be evaluated as  
\begin{align}\label{eq:gf}
\im G^>_{j} =&\im G^>_{j,0} = \langle \aop_j \adop_0 \rangle = \langle e^{-\im \Hop j\delta t} \aop e^{\im \Hop j\delta t} \adop \rangle \nonumber \\
=&\Zimp^{-1} \int\mea[\boldabar,\bolda] a_j^+ \abar_0^+ \gA[\boldabar,\bolda]; \\
-\im G^<_{j} =& -\im G^<_{j,0} = \langle \adop_0 \aop_j\rangle = \langle \adop e^{-\im \Hop j\delta t} \aop e^{\im \Hop j\delta t} \rangle \nonumber \\
=& \Zimp^{-1} \int\mea[\boldabar,\bolda] \abar_0^-  a_j^+ \gA[\boldabar,\bolda].
\end{align}
where the spin indices have been neglected.
In the zipup algorithm, this boils down to contracting a quasi-2D tensor network of size $3\times 8N$~\cite{ChenGuo2024b}. 
Assuming that the bond dimension of the GMPS representation of each $\gI_{\sigma}$ (referred to as the MPS-IF afterwards) is $\chi$, then the computational cost to build each MPS-IF is roughly $O(N^2\chi^3)$ using the partial IF algorithm~\cite{ChenGuo2024a} (which is improved to $O(N\chi^4)$ using a more efficient strategy to build the MPS-IF~\cite{GuoChen2024d}), and the computational cost to calculate one Green's function is roughly $O(N\chi^{3})$. 

When focusing on the NESS, unfortunately, the prefactor $N$ in the cost of GTEMPO (as well as most of its alternatives) is usually huge to overcome the transient dynamics, which significantly hinders the computational efficiency and necessitates more efficient approaches.

\section{The iGTEMPO Method} 
\gcc{As clear from the previous section, there are two major steps in the GTEMPO method: (1) building $\gK$ and $\gI_{\sigma}$ as GMPSs and (2) computing multi-time correlations based on them. In this section we will how to implement these two steps based on infinite GMPSs.

\subsection{The time-translationally invariant approach to build $\gK$ and $\gI_{\sigma}$ as infinite GMPSs}\label{app:TTIIF}

To motivate the infinite GMPS representation of the Feynman-Vernon IF, we first note that each \gcc{discretized} hybridization function $\Delta_{j, k}^{\zeta\zeta'}$ in Eq.(\ref{eq:gI}) is a single-variate function of the time difference $j-k$ only, e.g., $\Delta_{j,k}^{\zeta \zeta'} = \eta_{j-k}^{\zeta \zeta'}$. Therefore, $\gF_{\sigma}$, the exponent of $\gI_{\sigma}$, is invariant under any shift of its time step indices, which closely resembles an infinite-range many-body Hamiltonian. This time-translationally invariant (TTI) property of $\gF_{\sigma}$ and thus $\gI_{\sigma}$ has a profound impact: they can be represented as infinite GMPSs where only the site tensors within a single time step are independent. 

\begin{figure*}
  \includegraphics[width=2\columnwidth]{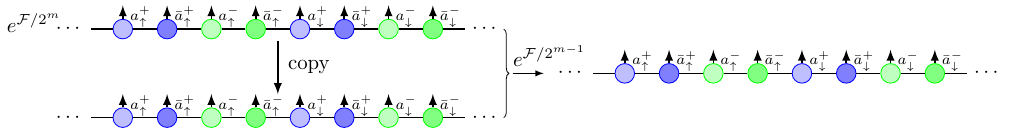} 
  \caption{Schematic demonstration of the time-translationally invariant approach to build $\gI_{\sigma}$ as an infinite Grassmann MPS using $m$ infinite GMPS multiplications (only the first step is shown here, since the rest $m-1$ steps are exactly the same as the first step), where compression is performed after each multiplication.
    }
    \label{fig:fig1S}
\end{figure*}

In the tensor network IF method~\cite{ThoennissAbanin2023b}, the Feynman-Vernon IF is built as a fermionic MPS in the Fock state basis using the Fishman-White algorithm~\cite{fishman2015-compression}. In the GTEMPO method, the Feynman-Vernon IF is built as a GMPS directly in the coherent state basis (which is the basis used for the analytical expression of the Feynman-Vernon IF) using the partial IF algorithm~\cite{ChenGuo2024a} by decomposing the IF into the product of a series of partial IFs, each with bond dimension $2$ only. However, neither approach respects the TTI property of the IF. In Ref.~\cite{GuoChen2024d}, the TTI property of the Feynman-Vernon IF is explicitly explored, which results in a very efficient algorithm (referred to as the TTI-IF algorithm) that not only respects the time-translational invariance, but also requires only a constant number of GMPS multiplications. The TTI-IF algorithm proposed in Ref.~\cite{GuoChen2024d} is used in the context of non-equilibrium real-time dynamics to build the finite GMPS representation of the IF, where the usage of open boundary GMPS is because the bare impurity dynamics part $\gK$ is not TTI. Nevertheless, the formalism of it can be readily used here to build the infinite GMPS representation of the IF. Here we will briefly review the major steps of the TTI-IF algorithm and stress the implementation-wise difference for the case of infinite GMPS.

We start from the discretized expression of the IF Eq.(\ref{eq:gI}).
In the finite case $N$ is determined by the total evolution time, i.e., $N = t / \delta t$ for total time $t$ and discrete time step size $\delta t$. 
In this work we directly focus on the steady state which is the infinite-time limit, and we will choose a large enough $N$ such that $|\Delta_{j,j+N}^{\zeta \zeta'}| \approx 0$ (in our numerical implementation we require the absolute value to be less than $10^{-6}$). 

In the next we denote 
\begin{align}\label{eq:gF2}
\gF_{\sigma}^{\zeta \zeta'} = -\sum_{j,k=1}^N\abar^{\zeta}_{\sigma,j}\Delta_{j,k}^{\zeta \zeta'}a^{\zeta'}_{\sigma,k}
\end{align}
and therefore $\gF_{\sigma} = \sum_{\zeta, \zeta'} \gF_{\sigma}^{\zeta \zeta'} $.
The TTI-IF algorithm contains two steps: (1) obtaining an efficient GMPS representation of $e^{\delta \gF_{\sigma}}$ with $\delta = 1/2^m$ a small positive number; (2) multiplying $2^m$ $e^{\delta \gF}$s together to obtain $\gI_{\sigma}$. 

For the first step, we note that
as each discretized hybridization function $\Delta_{j,k}^{\zeta \zeta'}$ is actually a single-variate function of the time difference $j-k$, $\gF_{\sigma}^{\zeta \zeta'}$ closely resembles a one-dimension translationally invariant many-body Hamiltonian. Therefore one could use the idea of constructing efficient infinite matrix product operator representation of translationally invariant Hamiltonians to build a compact GMPS representation of each $\gF_{\sigma}^{\zeta \zeta'}$. Now we explicitly denote
\begin{align}
\Delta_{j,k}^{\zeta \zeta'} = \eta_{j-k}^{\zeta \zeta'}
\end{align}
to stress the TTI property of the discretized hybridization function. We can use the Prony algorithm~\cite{marple2019digital} to find an optimal expansion of $\eta_{j-k}^{\zeta \zeta'}$ as the summation of $n$ exponential functions as
\begin{align}\label{eq:prony}
\eta_x^{\zeta \zeta'} \approx \sum_{l=1}^n \alpha_l \lambda_l^{|x|},
\end{align}
for both $x>0$ (for terms in Eq.(\ref{eq:gF2}) with $j>k$) and $x<0$ (for terms in Eq.(\ref{eq:gF2}) with $j < k$). Once the optimal values of $\alpha_l $ and $\lambda_l$ are obtained, we can construct $\gF_{\sigma}^{\zeta \zeta'}$ as a TTI GMPS whose site tensors are:
\begin{align}\label{eq:TTI1}
\left[\begin{array}{ccccccccc} 1 & \alpha_1 a^{\zeta'}_{\sigma} & \cdots & \alpha_n a^{\zeta'}_{\sigma} & -\bar{\alpha}_1 \abar^{\zeta}_{\sigma} & \cdots & -\bar{\alpha}_n \abar^{\zeta}_{\sigma} & \eta_0^{\zeta \zeta'} a^{\zeta'}_{\sigma}\abar^{\zeta}_{\sigma} \\
                 0 & \lambda_1 & \cdots & 0 & 0 & \cdots & 0 & \lambda_1 \abar^{\zeta}_{\sigma} \\  
                 \vdots & \vdots & \cdots & \vdots & \vdots & \cdots & \vdots & \vdots \\  
                 0 & 0 & \cdots & \lambda_n & 0 & \cdots & 0 & \lambda_n \abar^{\zeta}_{\sigma} \\  
                 0 & 0 & \cdots & 0 & \bar{\lambda}_1 & \cdots & 0 & \bar{\lambda}_1 a^{\zeta'}_{\sigma} \\
                 \vdots & \vdots & \cdots & \vdots & \vdots & \cdots & \vdots & \vdots \\   
                 0 & 0 & \cdots & 0 & 0 & \cdots & \bar{\lambda}_n & \bar{\lambda}_n a^{\zeta'}_{\sigma} \\  
                 0 & 0 & \cdots & 0 & 0 & \cdots & 0 & 1 \\  
\end{array}\right],
\end{align}
where $\alpha_l$ and $\lambda_l$ correspond to the expansion of $\eta_x^{\zeta \zeta'}$ for $1\leq x\leq N$ in Eq.(\ref{eq:prony}), while $\bar{\alpha}_l$ and $\bar{\lambda}_l$ correspond to the expansion of $\eta_x^{\zeta \zeta'}$ for $-N\leq x \leq -1$. 
For the iGTEMPO method, the only formal difference with the finite case considered in Ref.~\cite{GuoChen2024d} is that Eq.(\ref{eq:TTI1}) is understood as the site tensor of an infinite GMPS, while in Ref.~\cite{GuoChen2024d} it is understood as the site tensor of an open boundary GMPS. Here we also notice that Eq.(\ref{eq:TTI1}) contains two Grassmann variables (GVs), $a^{\zeta'}_{\sigma}$ and $\abar^{\zeta}_{\sigma}$, therefore if one represents each GV with a site tensor (which is the case in our implementation), then Eq.(\ref{eq:TTI1}) represents a two-site tensor which needs to be split into two tensors in practice.
With the infinite GMPS representation of each $\gF_{\sigma}^{\zeta \zeta'}$, one can immediately obtain a first-order approximation of $e^{\delta \gF_{\sigma}^{\zeta \zeta'}}$, using the $\WI$ algorithm~\cite{ZaletelPollmann2015} for example, which is an infinite GMPS with bond dimension $2n+1$, with TTI site tensors:
\begin{widetext}
\begin{align}\label{eq:exp-delta-F}
\left[\begin{array}{ccccccccc} 1+\delta\eta_0^{\zeta \zeta'} a^{\zeta'}_{\sigma}\abar^{\zeta}_{\sigma} & \sqrt{\delta}\alpha_1 a^{\zeta'}_{\sigma} & \cdots & \sqrt{\delta}\alpha_n a^{\zeta'}_{\sigma} & -\sqrt{\delta}\bar{\alpha}_1 \abar^{\zeta}_{\sigma} & \cdots & -\sqrt{\delta}\bar{\alpha}_n \abar^{\zeta}_{\sigma} \\
                 \sqrt{\delta}\lambda_1 \abar^{\zeta}_{\sigma} & \lambda_1 & \cdots & 0 & 0 & \cdots & 0  \\  
                 \vdots & \vdots & \cdots & \vdots & \vdots & \cdots & \vdots \\  
                 \sqrt{\delta}\lambda_n \abar^{\zeta}_{\sigma} & 0 & \cdots & \lambda_n & 0 & \cdots & 0 \\  
                 \sqrt{\delta}\bar{\lambda}_1 a^{\zeta'}_{\sigma} & 0 & \cdots & 0 & \bar{\lambda}_1 & \cdots & 0 \\
                 \vdots & \cdots & \vdots & \vdots & \cdots & \vdots & \vdots \\   
                 \sqrt{\delta}\bar{\lambda}_n a^{\zeta'}_{\sigma}  & 0 & \cdots & 0 & 0 & \cdots & \bar{\lambda}_n \\  
\end{array}\right].
\end{align}
\end{widetext}
In practice we use the slightly more sophisticated $\WII$ method instead, which results in an infinite GMPS with the same bond dimension but is a more accurate first-order approximation~\cite{ZaletelPollmann2015}.

Once we have obtained efficient infinite GMPS representations of the four $e^{\delta \gF_{\sigma}^{\zeta \zeta'}}$s (for the four combinations of $\zeta\zeta'$), we can multiply them together using infinite GMPS multiplications (which can be implemented in exactly the same way as finite GMPS multiplications~\cite{ChenGuo2024a}) to obtain $e^{\delta \gF_{\sigma}}$ as an infinite GMPS. During this process, infinite GMPS compression is done to compress the resulting infinite GMPS into a given bond dimension $\chi$ (the details for infinite GMPS compression will be discussed later). 
With $e^{\delta \gF_{\sigma}}$, Ref.~\cite{GuoChen2024d} introduces an extremely efficient algorithm to build $\gI_{\sigma}$ as a GMPS using only $m$ GMPS multiplications, which can be directly used here except that we deal with infinite GMPSs instead: in the $i$th step we multiply $e^{\gF_{\sigma}/2^{m-i+1}}$ with itself. This TTI-IF algorithm for building the infinite GMPS representation of $\gI_{\sigma}$ is schematically shown in Fig.~\ref{fig:fig1S}.

There are two additional sources of error in the TTI-IF algorithm on top of the time discretization error and the MPS bond truncation error (the latter two are the only sources of errors in the partial IF algorithm), namely the error occurred in the Prony algorithm and the error in the first order approximation of $e^{\delta \gF_{\sigma}}$. 
The Prony algorithm is a well studied algorithm in signal processing and can often converge with an $n$ that only scales logarithmically with $N$~\cite{croy2009-partial,zheng2009-numerical,hu2010-pade,VilkoviskiyAbanin2023}. 
The second source of error clearly decays exponentially with $m$ by the design of the algorithm.
In all the simulations of this work, we require the mean square error occurred in the Prony algorithm to be less than $10^{-5}$ and set $m=5$ (the same settings as in Ref.~\cite{GuoChen2024d}), which gives very accurate results for the model settings we have considered later in Sec.~\ref{sec:results}.

\begin{figure*}
  \includegraphics[width=\textwidth]{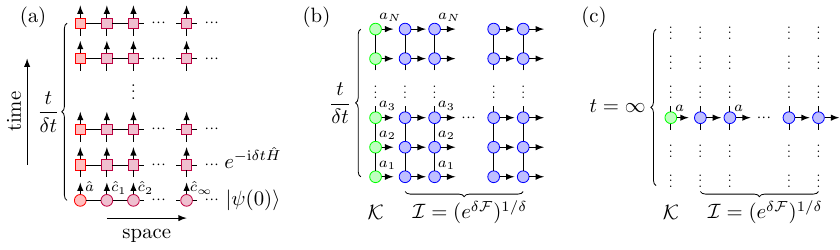} 
  \caption{
  \gcc{Schematic illustration of the essential calculations in (a) the conventional wave-function based MPS method, (b) the GTEMPO method and (c) the iGTEMPO method as the contraction of 2D tensor networks, where the bath and spin indices are neglected for briefness. The vertical axis is the temporal direction in all panels. The horizontal axis is the spatial direction in (a), whereas it has no particular meaning in (b,c). (a) The tensor network has size $N\times \infty$ ($N=t/\delta t$), where the bottom row is an MPS representation of the impurity-bath wave function and the rest rows are discretized time-evolutionary operators represented as matrix product operators (MPOs). The contraction is usually performed from bottom up using TEBD~\cite{Vidal2007}, TDVP~\cite{HaegemanVerstraete2011,VanderstraetenVerstraete2019} or standard MPO-MPS arithmetics~\cite{Schollwock2011}. (b) The tensor network has size $N\times 1/\delta$ with $\delta$ the ``time step'' size for discretizing $\gI$, where the left column is the GMPS representation of $\gK$ and the rest columns are multiplied together using finite GMPS multiplications~\cite{ChenGuo2024a} to obtain the GMPS representation of $\gI$. (c) Similar to (b), except that the temporal direction is extended to infinity and each column is represented as an infinite GMPS instead. The resulting tensor network has size $\infty \times 1/\delta$, and the information about the transient dynamics is lost.}
    }
    \label{fig:fig1}
\end{figure*}

In comparison to the Feynman-Vernon IF, the time-translational invariance of $\gK$ in Eq.(\ref{eq:gK}) is broken by 
the initial condition $\mel*{\bolda_{1}^+}{\rhoimp}{\bolda_{1}^-}$, which is the reason why open boundary GMPSs were used in the non-equilibrium setup~\cite{GuoChen2024d}. Crucially, in the steady state the memory of the initial state is completely lost, which means that this term can be neglected. The bulk terms of $\gK$, namely $\mel*{\bolda_{j+1}^+}{\Uimp}{\bolda_j^+} \mel*{\bolda_j^-}{\Uimp^{\dagger}}{\bolda_{j+1}^-}$, are TTI under the shift of the time step index $j$, whereas the boundary condition is taken at $j=\infty$. This is exactly the same situation met in the case of a one-dimensional MPS with \textit{infinite boundary condition}. Drawing this connection, we can represent $\gK$ as an infinite GMPS as well, and build it in a similar manner to that in the finite case but only perform the operations inside one unit cell. 


The compression of infinite GMPS is the only function that needs to be reimplemented when constructing $\gK$ and $\gI_{\sigma}$ as infinite GMPSs, compared to the finite case.
This can be done either deterministically using a series of singular value decompositions (SVDs) (\textit{SVD compression})~\cite{OrusVidal2008}, or iteratively such as using the infinite density matrix renormalization group (IDMRG) algorithm~\cite{McCulloch2008} or the variational uniform matrix product state (VUMPS) algorithm~\cite{StauberHaegeman2018}. We have implemented the first two approaches, and tested them in our numerical examples. As infinite MPS compression is a well-known technique that is independent from the main content of this work, we will not show the details of these techniques here (one could refer to the relevant papers cited above for the details of this technique).

\gcc{
To this end, we summarize the major difference between the conventional wave-function based MPS method, the GTEMPO method and the iGTEMPO method.
Roughly speaking, the major calculations in these methods can all boil down to contracting 2D tensor networks. However, the content and the sizes of the 2D tensor networks are very different, as illustrated in Fig.~\ref{fig:fig1}. We can clearly see that the sizes of the tensor networks involved in GTEMPO and iGTEMPO are significantly smaller than that in the conventional wave-function based MPS method, and that the difference between the GTEMPO and iGTEMPO methods completely reduces to the difference between finite and infinite MPS operations.}

\subsection{Calculating multi-time impurity correlations}\label{app:corr}

\begin{figure*}
  \includegraphics[width=2\columnwidth]{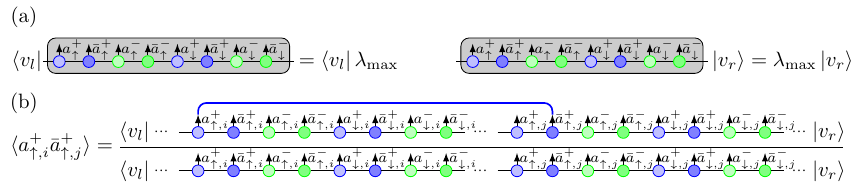} 
  \caption{Algorithm to calculate the Green's functions based on the infinite Grassmann MPS representation of the augmented density tensor, which can be done in two steps: (a) Computing the transfer matrix by integrating out the conjugate pairs of Grassmann variables in a unit cell, and then calculating the left and right dominant eigenstates of it; (b) Identifying a finite window from the infinite ADT depending on the Green's function to be calculated, and then evaluating the Green's function similar to the finite case, but with nontrivial left and right boundary vectors obtained from step (a). The single-orbital Anderson impurity model with $8$ Grassmann variables per unit cell is used for demonstration of the algorithm.
    }
    \label{fig:fig2S}
\end{figure*}

Once we have obtained the infinite GMPS representations of $\gK$ and $\gI_{\sigma}$, we can further multiply them together to obtain the ADT $\gA$ as an infinite GMPS (again $\gA$ is only calculated on the fly) and then compute multi-time impurity correlations based on it.
In GTEMPO, this is done by performing a left-to-right sweep, starting from and ending with trivial boundaries (e.g., the Grassmann vacuum $1$), during which the conjugate pairs of GVs are integrated out~\cite{ChenGuo2024a,ChenGuo2024b}. 

Based on the infinite GMPS representation of $\gA$, the Green's functions in Eq.(\ref{eq:gf}) (or any multi-time impurity correlations) can be calculated in two steps: (1) Obtaining the transfer matrix by integrating out each pair of GVs $a$ and $\abar$ in one unit cell with the measure $d \abar d a e^{-\abar a}$ (which boils down to contracting the two physical indices of the two site tensors corresponding to $a$ and $\abar$~\cite{ChenGuo2024a}, here we also note that this transfer matrix is completely different from the transfer matrix used when preparing an infinite MPS into the canonical form~\cite{OrusVidal2008}), and then calculating the dominant left and right eigenstates of it, denoted as $\langle v_l\vert $ and $\vert v_r\rangle $ respectively, with dominant eigenvalue denoted as $\lambda_{\max}$; (2) Identifying a finite window from the infinite GMPS representation of the ADT, and then evaluating the expectation value similar to the finite case, but using $\langle v_l\vert$ and $\vert v_r\rangle$ as the left and right boundaries instead of trivial boundaries. These two steps are schematically shown in Fig.~\ref{fig:fig2S}(a, b) respectively. 

}

\section{Numerical results}\label{sec:results}

In the next we benchmark the iGTEMPO method with concrete numerical examples, mostly in terms of accuracy, as its performance advantage compared to GTEMPO (GTEMPO is already very efficient for transport problems~\cite{ChenGuo2024a}) is essentially the difference between finite and infinite MPS algorithms.
For all our numerical simulations we will use a semi-circular bath spectrum density
\begin{equation}
  J(\omega) = \Gamma D\sqrt{1 - (\omega/D)^2} / 2\pi
\end{equation}
with $D=2$ and $\Gamma=0.1$ (we use $\Gamma$ as the unit), \gcc{which is frequently used in literatures to benchmark new impurity solvers~\cite{WolfSchollwock2014,WolfSchollwock2015,BertrandWaintal2019,ThoennissAbanin2023a}}. We have used IDMRG for infinite GMPS compression in the following numerical results as it is more efficient than SVD compression, whereas the comparison between these two compression algorithms can be found in the Appendix~\ref{app:compression}.

\subsection{Equilibrium Green's functions} 

\begin{figure}
  \includegraphics[width=\columnwidth]{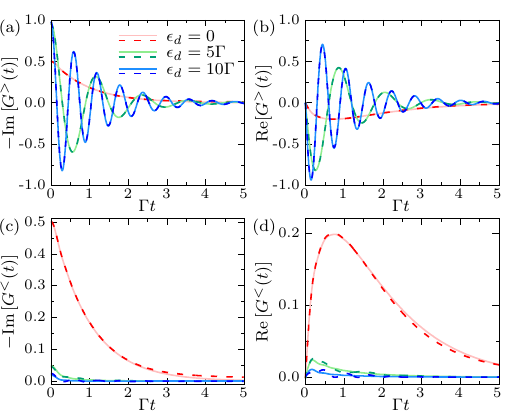} 
  \caption{(a) The real part and (b) the imaginary part of the equilibrium greater Green's function $G^>(t)$ as a function of $t$ in the noninteracting limit. (c) The real part and (d) the imaginary part of the equilibrium lesser Green's function $G^<(t)$ as a function of $t$. The red, green and blue dashed lines are iGTEMPO results for $\epsilon_d/\Gamma=0, 5, 10$ respectively, the solid lines with the same colors are the corresponding ED results. We have used $\Gamma\delta t=0.005$, $\chi=60$ and $\Gamma\beta=4$ in these simulations.
    }
    \label{fig:fig3}
\end{figure}

We first validate iGTEMPO in the noninteracting limit against exact solutions. 
In Fig.~\ref{fig:fig3}, we compare the equilibrium greater and lesser Green's functions calculated using iGTEMPO (with $\chi=60$ and $\Gamma \delta t = 0.005$) with exact diagonalization (ED) results calculated with $\delta\omega/\Gamma=0.005$ (the bath discretization is the only error in ED, and we have verified that our ED results have well converged against $\delta\omega$). 
We can see that the iGTEMPO results well agree with ED for different values of $\epsilon_d$, where the largest error is within $1\%$ (See Appendix.~\ref{app:compression} for the convergence of iGTEMPO results against the bond dimension).

\begin{figure}
  \includegraphics[width=\columnwidth]{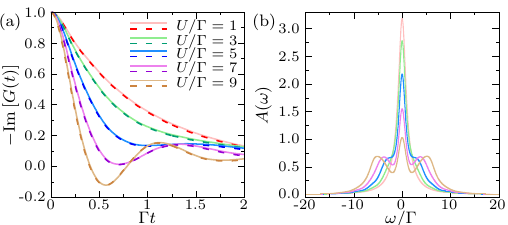} 
  \caption{(a) The imaginary part of the equilibrium retarded Green's function of single-orbital AIM as a function of $t$ for different values of $U$, where the solid and dashed lines are iGTEMPO and GTEMPO results respectively. For both iGTEMPO and GTEMPO we have used $\Gamma\delta t=0.005$, $\chi=60$ and $\Gamma\beta=4$. (b) The spectral function $A(\omega)$ as a function of frequency $\omega$, obtained by taking the Fourier transformation of $G(t)$.
    }
    \label{fig:fig4}
\end{figure}

Then we calculate the finite-temperature equilibrium retarded Green's function $G(t) = G^>(t) - G^<(t)$ of the single-orbital AIM for which there does not exist exact solutions. 
Nevertheless, the GTEMPO results by calculating $G(t-t_0)$ with $\Gamma t_0=8$ (to overcome the transient dynamics) are already very accurate, since the impurity and bath have well reached equilibrium after $t_0$ for the considered set of parameters~\cite{ChenGuo2024c}. 
In Fig.~\ref{fig:fig4}(a) we compared the iGTEMPO results with the GTEMPO results, for different values of $U/\Gamma$ ranging from $1$ to $9$. We can see very good matches between these two sets of results for all $U$s. In Fig.~\ref{fig:fig4}(b) we show the spectral function $A(\omega) = -{\rm Im}[G(\omega)]/\pi$ where $G(\omega)$ is the Fourier transformation of $G(t)$. To calculated $A(\omega)$ we have used linear prediction~\cite{BarthelWhite2009} to extend $G(t)$ to very large $t$ such that $|G(t)| < 10^{-6}$. To this end, we also note another advantage of (i)GTEMPO compared to both CTQMC and the wave-function based methods: once the $\gI_{\sigma}$s have been built as (infinite) GMPSs, they can be saved and used later with the $\gK$s for different impurity Hamiltonians (the cost of building $\gK$ is negligible) to compute the multi-time impurity correlations (as an example, the MPS-IFs for Fig.~\ref{fig:fig4} are only calculated once for all different $U$s). 

\subsection{Non-equilibrium steady state} 

Finally, we calculate the NESS of an impurity coupled to two baths with a voltage bias, an ideal application scenario for iGTEMPO. We consider the two baths at zero temperature with voltage bias $\mu_1 = -\mu_2 = V/2$, and compute the symmetrized steady-state particle current, defined as $\current=(\current_{\uparrow}^1 - \current_{\uparrow}^2)/2 = (\current_{\downarrow}^1 - \current_{\downarrow}^2)/2$, where $\current_{\sigma}^{\nu}$ denotes the particle current with spin $\sigma$ that flows from the $\nu$th bath into the impurity. In the (i)GTEMPO methods, the particle current can be calculated as a summation of single particle Green's functions~\cite{ChenGuo2024a}.

\begin{figure}
  \includegraphics[width=\columnwidth]{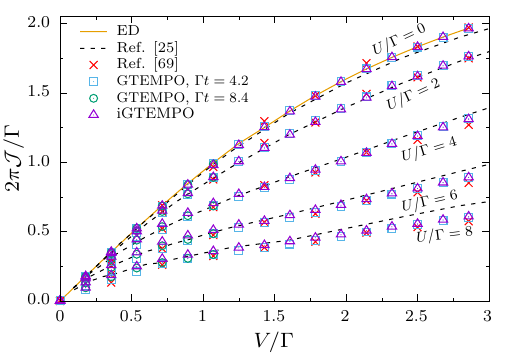} 
  \caption{The symmetrized steady-state particle current $\mathcal{J}$ in the non-equilibrium AIM as a function of the voltage bias $V$ for different values of $U$, calculated by ED (brown solid line for $U=0$), improved Quantum Monte Carlo~\cite{BertrandWaintal2019} (black dashed line), tensor network IF till $\Gamma t=4.2$ (with $\chi=32$ and $\Gamma \delta t=0.007$)~\cite{ThoennissAbanin2023b} (red x), GTEMPO till $\Gamma t=4.2$ (with $\chi=160$ and $\Gamma \delta t=0.007$)~\cite{ChenGuo2024a} (cyan square) and till $\Gamma t=8.4$ (with $\chi=160$ and $\Gamma \delta t=0.014$)~\cite{GuoChen2024d} (green circle), and iGTEMPO (purple triangle). For iGTEMPO we have used $\chi=60$ and $\Gamma \delta t=0.005$.
    }
    \label{fig:fig5}
\end{figure}

The steady-state particle current of this model has been calculated by an improved Quantum Monte Carlo (QMC) method~\cite{BertrandWaintal2019}, by the tensor network IF method till $\Gamma t=4.2$ (with $\chi=32$ and $\Gamma\delta t=0.007$)~\cite{ThoennissAbanin2023b}, by the GTEMPO method till $\Gamma t=4.2$ (with $\chi=160$, $\Gamma\delta t=0.007, 0.014$)~\cite{ChenGuo2024a} and till $\Gamma t=8.4$ (with $\chi=160$, $\Gamma\delta t=0.014$ for $V/\Gamma < 1.1$) using a more efficient method to build the MPS-IF~\cite{GuoChen2024d}.
In the noninteracting limit with $U=0$, the GTEMPO and iGTEMPO results both agree very well with the ED results. In the interacting case, the time required to reach the steady state seems to be larger for smaller $V$ and larger $U$, and it has been shown that the GTEMPO results have well converged for $V/\Gamma\geq 1$~\cite{GuoChen2024d}.
In Fig.~\ref{fig:fig5}, we show the steady-state particle current calculated by iGTEMPO, with comparisons to the existing calculations. We can see that the GTEMPO results calculated with $\Gamma t = 8.4$ agree fairly well with the QMC results except for the points with $V/\Gamma \approx 0.17$ and $U/\Gamma > 2$ (the GTEMPO results for this point may have not converged yet with $\Gamma t=8.4$). The iGTEMPO results also agree fairly well with the QMC results, except for $V/\Gamma \approx 0.17$ with $U/\Gamma > 2$ and for $V/\Gamma \approx 0.54, 0.71$ with $U/\Gamma \geq 4$, but it is not clear for now which set of results is more accurate at these points. 

\gcc{Here we note another advantage of iGTEMPO compared to GTEMPO besides that iGTEMPO directly aims for the steady state: in iGTEMPO one can easily use a smaller time step size $\delta t$ as the computational cost to build $\gK$ and $\gI_{\sigma}$ is essentially independent of the total number of time steps (the cost of calculating multi-time correlations, in comparison, will grow for smaller $\delta t$, as the finite window size will become effectively larger, nevertheless, the latter step can be easily parallelized as the calculations for different multi-time correlations are completely independent of each other). Therefore it is easier for one to reduce the time discretization error in iGTEMPO. In addition, in Fig.~\ref{fig:fig5} we have used a smaller bond dimension for iGTEMPO than that used for GTEMPO, but reached a similar level of precision, which may indicate that infinite GMPSs could be more expressive for the steady state than finite GMPSs. However, more numerical investigations are still required to confirm this point, which could be done in future studies.}

\section{Conclusion}

In summary, we have proposed an infinite Grassmann time-evolving matrix product operator (iGTEMPO) method which makes use of the infinite Grassmann matrix product state to represent the multi-time impurity steady state. 
Similar to GTEMPO, iGTEMPO directly obtains real-time Green’s functions without sampling noises, applicable for any temperature without the sign problem, and its computational cost does not scale with the number of baths.
Compared to GTEMPO or other non-sampling based methods, the infinite GMPS ansatz used in iGTEMPO is extremely compact: only the site tensors in a single time step is independent, and the total number of parameters only scale as $16\chi^2$ for the single-orbital Anderson impurity model.
The computational cost of iGTEMPO is essentially independent of the total evolution time, which roughly scales as $O(\chi^3)$ for the single-orbital AIM.
For the single-orbital AIM with the commonly used semi-circular spectrum in both the equilibrium (one bath) and non-equilibrium (two baths) setups, we show that with $\chi = 60$ we can already obtain results with comparable precision to existing calculations. The iGTEMPO method is ideal for studying steady-state quantum transport problems, and can be readily used as a real-time impurity solver in the dynamical mean field theory (DMFT)~\cite{GullWerner2011} or non-equilibrium DMFT~\cite{AokiWerner2014}.
\gcc{Generalizing the central techniques developed in this work for bosonic impurity problems is straightforward (which could be done by following Ref.~\cite{GuoChen2024d} in the finite case), where standard infinite MPSs can be used instead of infinite GMPSs. We also note that a very different strategy has already been proposed for bosonic impurity problems which integrates infinite MPS techniques into the time-evolving matrix product operator method~\cite{LinkStrunz2024}.}

\begin{acknowledgments}
We thank Shiju Ran for helpful discussions on infinite MPS algorithms.
This work is supported by National Natural Science Foundation of China under Grant No. 12104328. C. G. is supported by the Open Research Fund from State Key Laboratory of High Performance Computing of China (Grant No. 202201-00).
\end{acknowledgments}

\appendix

\section{Infinite GMPS compression}\label{app:compression}

\begin{figure*}
  \includegraphics[width=2\columnwidth]{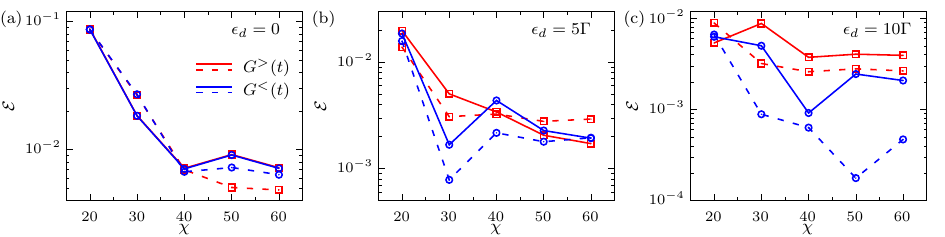} 
  \caption{Average error between the iGTEMPO results and ED results for (a) $\epsilon_d/\Gamma=0$, (b) $\epsilon_d/\Gamma=5$ and (c) $\epsilon_d/\Gamma=10$. The red solid and dashed lines represent the iGTEMPO results for the equilibrium greater Green's functions calculated using SVD compression and IDMRG compression respectively. The blue solid and dashed lines represent the iGTEMPO results for the equilibrium lesser Green's functions calculated using SVD compression and IDMRG compression respectively. We have used $\Gamma \delta t=0.005$ and $\Gamma\beta=4$ in all these simulations.
    }
    \label{fig:fig3S}
\end{figure*}

The multiplication of two MPSs with bond dimensions $\chi_1$ and $\chi_2$ will result in an MPS with bond dimension $\chi_1\chi_2$. In practice one sets a maximum bond dimension $\chi$ and compress the resulting MPS into bond dimension $\chi$ to maintain the computational cost~\cite{Schollwock2011}. The compression can be done exactly (in principle) and efficiently if the MPS has an exact canonical form, which is the case for both the open boundary MPS and the infinite boundary MPS. 
When performing compression of an infinite GMPS, the GMPS can be treated as a standard infinite MPS~\cite{ChenGuo2024a}. There already exists a number of algorithms to compress an infinite MPS, the representative ones include the deterministic SVD compression by performing a full left-to-right and then right-to-left sweep~\cite{OrusVidal2008}, and the iterative ones such as using the infinite density matrix renormalization group (IDMRG) algorithm~\cite{McCulloch2008}, or the variational uniform MPS (VUMPS) algorithm~\cite{StauberHaegeman2018}, with one-to-one correspondence to their open boundary MPS counterparts. In this work, we have considered two approaches, the SVD compression and the IDMRG compression. The advantage of the first approach is that it is deterministic and in principle leads to the optimal canonical form. However, in the TTI-IF algorithm as described in Sec.~\ref{app:TTIIF}, one needs to multiply two same infinite GMPS with bond dimension $\chi$, as a result the resulting infinite GMPS has a bond dimension $\chi^2$, and the cost of the SVD compression will scale as $O(\chi^6)$ as similar to the finite case~\cite{GuoChen2024d} (but without the prefactor $N$ for the total evolution time). Moreover, one needs to perform the inversion of the singular matrix (See Ref.~\cite{OrusVidal2008} for details), which is numerical unstable if the conditional number of the singular matrix is too large (which is often the case, unfortunately). In comparison, the IDMRG algorithm is iterative (which means that one may be trapped in non-optimal infinite GMPS approximations), while its computational cost for our problem only scales as $O(\chi^4)$ (in the iterative compression scheme, the multiplication of the infinite GMPSs are only computed on the fly to reduce the memory usage and the computational cost). In this work we use the single-site IDMRG algorithm following the implementation in the package MPSKit.jl~\cite{MPSKit}. When using the single-site IDMRG algorithm for our $Z_2$-symmetric MPS~\cite{ChenGuo2024a}, the virtual space will be fixed from the beginning. Since we only have two symmetry sectors $0$ and $1$, we initialize each symmetry sector with size $\chi/2$ and find that this is a good choice in practice. We typically use $10000$ to $30000$ IDMRG sweeps in the numerical simulations of the main text, since IDMRG essentially uses the power method to find the fixed point of the environment which generally converges quite slowly. 
We observe that the IDMRG compression is almost never trapped in very bad solutions as long as the bond dimension is large enough. In the future, one may use the more recent VUMPS algorithm for infinite GMPS compression, which may achieve faster convergences.

In Fig.~\ref{fig:fig3S}
we show the convergence of the iGTEMPO calculations against the increase of the bond dimension, and compare the accuracy of the two approaches (SVD compression and IDMRG) used for infinite GMPS compression in the noninteracting case. We use the average error $\mathcal{E}$, defined as $\mathcal{E}(\vec{x}, \vec{y}) = \sqrt{||\vec{x} - \vec{y} ||^2 / L}$ for two vectors $\vec{x}$ and $\vec{y}$ of length $L$, to quantify the derivation between the iGTEMPO results and the exact solutions calculated by exact diagonalization (ED). For SVD compression we have set the tolerance of the eigensolver (used to calculate the dominate eigenstates of the transfer matrix) to be $10^{-14}$, and for IDMRG we have used $20000$ sweeps.
We can see that in both approaches the average errors can be brought down to less than $1\%$ with $\chi=60$ for very different values of $\epsilon_d$, and the IDMRG results for $\epsilon_d/\Gamma=0, 10$ are slightly more accurate.


%

\end{document}